\documentclass{article}
\usepackage{amsmath}
\usepackage{graphicx}

\begin{document}

\title{Logarithmic Integrals of Airy Functions}
\author{Bernard J. Laurenzi \\
Department of Chemistry\\
UAlbany, The State University of New York\\
1400 Washington Ave., Albany N.Y. 12222}
\date{October 29, 2012}
\maketitle

\begin{abstract}
Integrals arising in the Thomas-Fermi (TF) theory of atomic structure and
which contain logarithms of the Airy functions have been expressed in terms
of the incomplete Bell polynomials. \ In keeping with the spirit of TF
theory closed forms for these integrals are sought.
\end{abstract}

\section{Introduction}

In the theory of atomic structure due to Thomas and Fermi \cite{tf}
integrals arise \cite{Laur} which contain the Airy function $Ai(x)$ and its
derivative $Ai^{\prime }(x)$ \cite{Airy}$.$ \ A typical example of these
integrals being%
\begin{equation*}
I=\int_{0}^{\infty }A^{2}(x)\ln [A(x)]\,dx,
\end{equation*}%
where $A(x)$ is given by 
\begin{equation*}
A(x)=\frac{Ai^{\prime }(x)}{Ai^{\prime }(0)}.
\end{equation*}%
The value of the integral (to 8 figures) obtained by numerical methods is
-0.26363171. \ However, it would be useful and within the spirit of Thomas
Fermi (TF) theory if an exact or at least an analytic expression for this
integral and others like it could be obtained. \ With this in mind, this
paper presents methods which attempt to move towards that goal. \ In
anticipation of the work which follows we set%
\begin{equation*}
z=1-A(x)
\end{equation*}%
and rewrite the integral above as 
\begin{equation*}
I=\int_{0}^{1}(1-z)^{2}\ln (1-z)\frac{dx}{dz}dz.
\end{equation*}%
Integration by parts gives 
\begin{equation}
I=\int_{0}^{1}(1-z)[1+2(1-z)\ln (1-z)]\,x(z)\,dz.  \label{eq1}
\end{equation}%
Where we see that $x(z)$ is the inverse function for $A$ with argument $1-z.$
\ The latter function can be obtained using the Lagrange expansion formula %
\cite{Math} i.e.%
\begin{equation}
x(z)=\sum_{k=1}^{\infty }\frac{a_{k}}{k!}\,(\sqrt{z})^{k},  \label{eq2}
\end{equation}%
with the coefficients $a_{k}$ given by 
\begin{equation*}
a_{k}=\left[ \frac{d^{\,k-1}}{d\,x^{\,k-1}}\left( \frac{1}{\sqrt{u(x)}}%
\right) ^{k}\right] _{x=0},
\end{equation*}%
where%
\begin{equation*}
u(x)=\left[ 1-A(x)\right] /x^{2}.
\end{equation*}%
Using (2), the integrals on the right hand side of (1) are elementary and
the integral $I$ can be written as%
\begin{equation*}
I=8\sum_{k=1}^{\infty }\frac{a_{k}}{k!}\,\frac{\left[ \psi (3)-\psi (k/2+3)%
\right] }{(k+2)(k+4)},
\end{equation*}%
where $\psi (k)$ is the Psi (digamma) function \cite{Psi}. \ 

\subsection{Evaluation of the $a_{k}$}

The derivatives within $a_{k}$ can in turn be given by the Fa\`{a} di Bruno
formula for the (k-1)$^{st}$ derivative of a composite function \cite{Faa}
as expressed in terms of the incomplete Bell polynomials $%
B_{k,\,n}(x_{1,}x_{2,}\ldots x_{k-n+1})$ \cite{InBell} i.e.%
\begin{equation*}
a_{k}=\sum_{p=1}^{k-1}\left[ \frac{d\,^{p}u^{-k/2}}{d\,u^{\,p}}\right]
_{x=0}B_{k-1,\,p}(u^{(1)}(0),u^{(2)}(0),\ldots ,u^{(k-p)}(0)),
\end{equation*}%
where 
\begin{equation*}
u^{(i)}(0)=\left[ \frac{d\,^{i}u(x)}{dx^{i}}\right] _{x=0}.
\end{equation*}%
The expression for $a_{k}$ can be quickly reduced to terms involving the
Gamma function $\Gamma (k)$ \cite{Gamma} to get 
\begin{equation*}
a_{k}=\tfrac{1}{u(0)^{k/2}}\sum_{p=1}^{k-1}\left( \tfrac{-1}{u(0)}\right)
^{p}\tfrac{\Gamma (k/2+p)}{\Gamma (k/2)}B_{k-1,\,p}(u^{(1)}(0),u^{(2)}(0),%
\ldots ,u^{(k-p)}(0)).
\end{equation*}%
The derivatives of the function $u(x)$ can easily be obtained since the
power series for $Ai^{\prime }(x)$ is well known \cite{Vallee} i.e.%
\begin{equation*}
Ai^{\prime }(x)/Ai^{\prime }(0)=1-\tfrac{3^{2/3}\Gamma (1/3)}{\pi }%
\sum_{n=0}^{\infty }\tfrac{\Gamma (\frac{n+4}{3})}{(n+2)!}\sin (\tfrac{2}{3}%
(n+4)\,\pi )\,(3^{1/3}x)^{n}.
\end{equation*}%
Using the latter expression we get 
\begin{eqnarray*}
u(0) &=&\tfrac{\pi }{3^{5/6}\Gamma (2/3)^{2}}, \\
u^{(i)}(0) &=&\left[ \tfrac{2}{3^{5/6}\,\Gamma (2/3)}\right] \,{\small 3}%
^{i\,/3}\,\tfrac{\Gamma (\frac{i+1}{3})\sin (\frac{2}{3}[i+1]\pi )}{(i+2)}.
\end{eqnarray*}%
Finally the integral is given by%
\begin{equation}
I=\tfrac{8}{3^{1/3}}\sum_{k=1}^{\infty }\,\tfrac{\left[ \psi (3)\,-\,\psi
(k/2+3)\right] }{(k+2)(k+4)}S_{k},  \label{eq3}
\end{equation}%
where $S_{k}$ is defined as%
\begin{equation}
S_{k}=\tfrac{\left( \frac{3^{3/4}\,\Gamma (2/3)}{\sqrt{\pi }}\right) ^{k}}{k!%
}\sum_{p=0}^{k-1}\left( \tfrac{-2\,\Gamma (2/3)}{\pi }\right) ^{p}\tfrac{%
\Gamma (k/2+p)}{\Gamma (k/2)}B_{k-1,\,p}(\widehat{u}^{(1)}(0),\widehat{u}%
^{(2)}(0),\ldots ,\widehat{u}^{(k-p)}(0)),  \label{eq4}
\end{equation}%
and the reduced quantities $\widehat{u}^{(i)}(0)$ are given by

\begin{equation*}
\widehat{u}^{(i)}(0)=\tfrac{{\small \Gamma \,(}\tfrac{i+1}{3}{\small )}\sin (%
\frac{2}{3}[i\,+1]\pi )}{(i+2)}.
\end{equation*}%
In obtaining the expression in (3,4) we have used the homogeneous scaling
properties \cite{Brauchart} of the incomplete Bell polynomials i.e.%
\begin{eqnarray*}
\alpha ^{n}\,B_{k,\,n}(x_{1,}x_{2,}\ldots ) &=&B_{k,\,n}(\alpha x_{1},\alpha
x_{2},\ldots ), \\
\alpha ^{k}\,B_{k,\,n}(x_{1,}x_{2,}\ldots ) &=&\,B_{k,\,n}(\alpha
x_{1,}\alpha ^{2}x_{2,}\ldots ),
\end{eqnarray*}%
to rewrite the Bell polynomials in terms of the reduced quantities $\widehat{%
u}^{(i)}(0).$

\subsection{The Rate of Convergence of the $I$ Integral}

The infinite sum representation for the integral $I$ in (3) is a slowly
varying function of $k$. \ For example, the value of $I$ given by a partial
sum containing the first ten terms is only 91 \% of the value obtained
by\thinspace numerical evaluation of the integral. Although a complete
analysis (or estimate) of the convergence behavior of this sum does not
appear to be possible, it is likely that (3) is an asymptotic series
representation of the integral $I$.\ \ Nevertheless, in an attempt to
increase the rate of convergence i.e. accelerate the assumed convergence of
the sum, we note that the magnitudes (Figure 1) of the summands $I_{k}$%
\begin{equation*}
I_{k}=\tfrac{8}{3^{1/3}}\tfrac{\left[ \psi (3)\,-\,\psi (k/2+3)\right] }{%
(k+2)(k+4)}S_{k},
\end{equation*}%
in (3) are small for values of $k$ greater than 10.

\begin{figure}[h!]
  \caption{$I_k$ vs. $k$}
  \centering
    \includegraphics[width=0.5\textwidth]{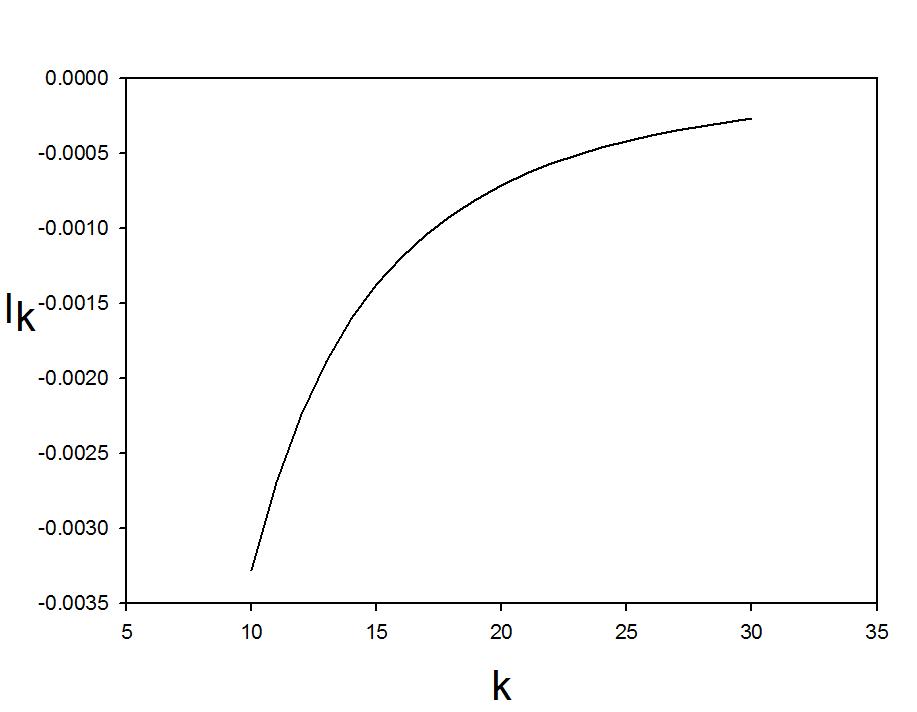}
\end{figure}

\bigskip\ \ Furthermore, when $S_{k}$ is plotted (Figure 2) versus $k$ for
values of $k\geq 10,$ it can be seen to vary like powers of $1/k$ i.e.%
\begin{equation*}
S_{k}=a/k+b/k^{2}+c/k^{3},
\end{equation*}%
where regression analysis yields $a$ = 0.751653834, $b$ = 2.25325549, $c$ =
-6.815672901.

\begin{figure}[h!]
  \caption{$S_k$ vs. $k$}
  \centering
    \includegraphics[width=0.5\textwidth]{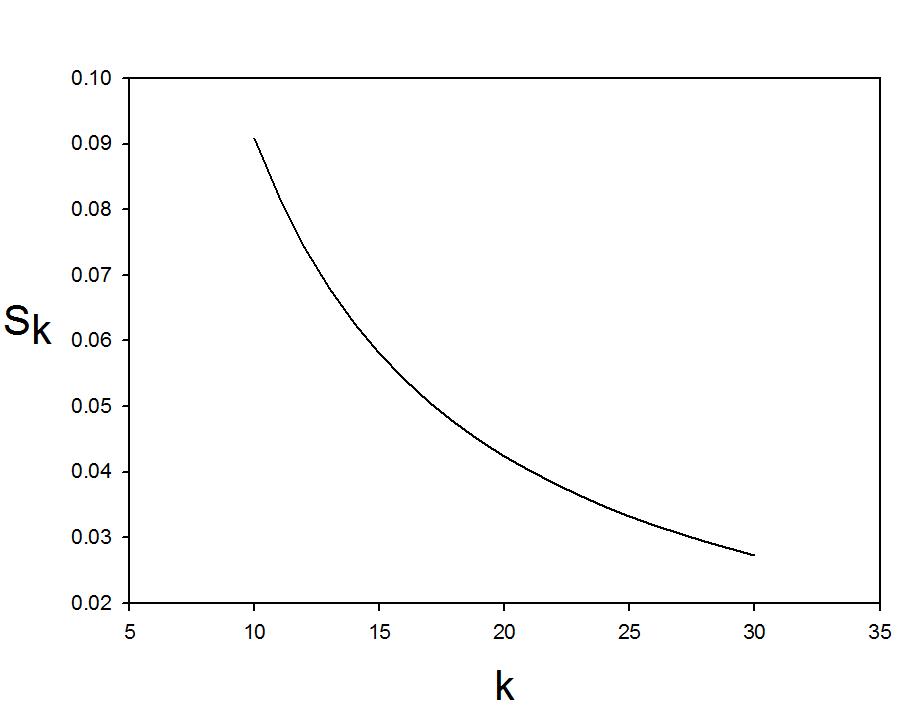}
\end{figure}
Using the curve fitted
expression for $S_{k}$ the expression for $I$ becomes%
\begin{equation*}
I=\tfrac{8}{3^{1/3}}\sum_{k=1}^{10}\,\tfrac{\left[ \psi (3)\,-\,\psi (k/2+3)%
\right] }{(k+2)(k+4)}S_{k}+I_{remainder},
\end{equation*}%
where 
\begin{eqnarray}
I_{remainder} &=&\tfrac{8}{3^{1/3}}{\Large [}a\left\{ \sigma
_{1}-\sum_{k=1}^{10}\tfrac{\left[ \psi (3)\,-\,\psi (k/2+3)\right] }{%
k\;(k+2)\;(k+4)}\right\} +b\left\{ \sigma _{2}-\sum_{k=1}^{10}\tfrac{\left[
\psi (3)\,-\,\psi (k/2+3)\right] }{k^{2}\;(k+2)\;(k+4)}\right\}  \label{eq5}
\\
&&+c\left\{ \sigma _{3}-\sum_{k=1}^{10}\tfrac{\left[ \psi (3)\,-\,\psi
(k/2+3)\right] }{k^{3}\;(k+2)\;(k+4)}\right\} {\Large ]}.  \notag
\end{eqnarray}%
The sums $\sigma _{i\text{ }}$appearing in (5) can be calculated in terms of
known \cite{sums}, closed forms i.e. 
\begin{equation*}
\sigma _{1}=\sum_{k=1}^{\infty }\tfrac{\left[ \psi (3)\,-\,\psi (k/2+3)%
\right] }{k\,(k+2)(k+4)\,}=-\tfrac{89}{576}+\tfrac{1}{6}\ln {\small (2)},
\end{equation*}%
\begin{equation*}
\sigma _{2}=\sum_{k=1}^{\infty }\tfrac{\left[ \psi (3)\,-\,\psi (k/2+3)%
\right] }{k^{2}\,(k+2)(k+4)\,}=\tfrac{349}{1152}-\tfrac{11}{16}\zeta (3)-%
\tfrac{1}{6}\ln {\small (2)},
\end{equation*}%
\begin{equation*}
\sigma _{3}=\sum_{k=1}^{\infty }\tfrac{\left[ \psi (3)\,-\,\psi (k/2+3)%
\right] }{\,k^{3}\,(k+2)(k+4)}=
\end{equation*}%
\begin{equation*}
=-\tfrac{2423}{9216}+\tfrac{5\pi ^{2}}{384}-\tfrac{11\pi ^{4}}{2880}-\tfrac{%
\pi ^{2}\ln ^{2}(2)}{96}+\tfrac{5\ln (2)}{48}+\tfrac{\ln ^{4}(2)}{96}%
+\left\{ \tfrac{7\ln (2)}{32}+\tfrac{33}{256}\right\} \zeta (3)+\tfrac{1}{4}%
Li_{4}{\small (1/2)},
\end{equation*}%
where $Li_{4}$ is the 4$^{th}$ order polylogarithm function \cite{Lewin} and 
$\zeta (z)$ is the Riemann zeta function. \ The value of the sums given
above are small with magnitudes -0.0389893588,\ -0.0191766714 ,
-0.0146522682 respectively. \ Using these values $I_{remainder}=$
-0.004280449344 yielding a value of -0.2637166702 for $I.$ \ The latter
value of $I$ having an error of 0.03 \%, this procedure is seen to produce
the best estimate for an ``analytic'' representation the integral thus far.

\ In order to make further progress in obtaining a closed form expression
for integrals of the kind sought here, closed form sums of the type%
\begin{equation}
\left( \tfrac{1}{k!}\right) \sum_{p=0}^{k-1}(k/2)_{p}\,B_{k-1,\,p}(\widehat{%
u}^{(1)}(0),\widehat{u}^{(2)}(0),\ldots ,\widehat{u}^{(k-p)}(0)),
\label{eq6}
\end{equation}%
where $(a)_{n}$ is the Pochhammer polynomial symbol and the $\widehat{u}%
^{(i)}(0)$ terms have been given above, must be found. \ Expressions for
closed form sums similar (6) have been given by Mihoubi \cite{mihoubi} and
Brauchart \cite{Brauchart} albeit with simpler arguments of the Bell
polynomials. \

\end{document}